\documentclass{article}%
\usepackage[top=3cm, bottom=3cm, left=3cm, right=3cm]{geometry}
\usepackage{amsmath}
\usepackage{amsfonts}
\usepackage{amssymb}
\usepackage{graphicx}%
\begin{document}

\title{Spontaneous breaking of chiral symmetry, and eventually of parity, in a
$\sigma$-model with two Mexican hats}
\author{Francesco Giacosa\\\emph{Institute for Theoretical Physics, }\\\emph{Johann Wolfgang Goethe University, }\\\emph{Max von Laue--Str. 1, 60438 Frankfurt am Main, Germany}}
\maketitle

\begin{abstract}
A $\sigma$-model with two linked Mexican hats is discussed. This scenario
could be realized in low-energy QCD when the ground state and the first
excited (pseudo)scalar mesons are included, and where not only in the subspace
of the ground states, but also in that of the first excited states, a Mexican
hat potential is present. This possibility can change some basic features of a
low-energy hadronic theory of QCD. It is also shown that spontaneous breaking
of parity can occur in the vacuum for some parameter choice of the model.

\end{abstract}


\bigskip

\section{Introduction}

The `Mexican hat' potential allows for a simple and intuitive description of
the phenomenon of spontaneous symmetry breaking. For this reason it has been
widely used -in a variety of versions- in both condensed matter and hadron
physics, see for instance Ref. \cite{zee} and refs. therein.

In the context of Quantum Chromodynamics (QCD) nearly massless $N_{f}^{2}-1$
(3 pions in the case $N_{f}=2$, where $N_{f}$ is the number of light quark
flavors) emerge as (quasi) pseudoscalar Goldstone bosons as a consequence of
spontaneous breaking of chiral symmetry: $U_{R}(N_{f})\times U_{L}%
(N_{f})\rightarrow SU_{V}(N_{f}).$ In the context of a linear $\sigma$-model
this spontaneous breaking is induced by a negative squared mass of the scalar
and pseudoscalar mesons. This feature is responsible for the typical Mexican
hat form of the mesonic potential.

In this work, beyond the ground state (pseudo)scalar mesons, we also consider
the first excited (pseudo)scalar states and we investigate the case in which
also in this sector a negative squared mass is present. As we shall argue, for
some parameter choice this possibility cannot be excluded and leads to a more
complicated scenario, in which ground-state and first-excited scalar and
pseudoscalar mesons mix. Moreover, for some parameter choice it is possible
that also one neutral pseudoscalar pionic field condenses, thus realizing a
spontaneous symmetry breaking of parity.

The paper is organized as it follows: we first briefly review the properties
of the Mexican hat potential and its emergence from an hadronic model of QCD.
We then turn to the case of two linked Mexican hats and discuss the
consequences of this assumptions. First, the parameter range in which only
spontaneous breaking of chiral symmetry take place is studied. Then, the
parameter range in which also spontaneous breaking of parity occurs is
investigated. In the end, the conclusions are briefly outlined.

\section{Mexican hat }

In its simplest form the Mexican hat potential is written in terms of two real
scalar fields $\sigma$ and $\pi$:%

\begin{equation}
V_{\text{MH}}=\frac{\lambda}{4}\left(  \sigma^{2}+\pi^{2}-F^{2}\right)
^{2}=\frac{\lambda}{4}\left(  \varphi^{\ast}\varphi-F^{2}\right)  ^{2}%
\begin{array}
[c]{c}%
,
\end{array}
\label{sigma}%
\end{equation}
where in the last passage the complex scalar field $\varphi=\sigma+i\pi$ has
been introduced. The requirement $\lambda\geq0$ ensures that the potential is
bounded from below. Let us assume that -as in QCD, see below- $\sigma$
represents a scalar field ($\sigma\equiv\sigma(t,\mathbf{x})\rightarrow
\sigma(t,-\mathbf{x})$ under parity transformation $P$) while $\pi$ represents
a pseudoscalar field ($\pi\equiv\pi(t,\mathbf{x})\rightarrow-\pi
(t,-\mathbf{x})$ under $P$). Note, the quadratic (mass) term of the Mexican
hat potential reads $-\frac{\lambda}{2}F^{2}\varphi^{\ast}\varphi,$ i.e. it
has a negative coefficient as long as $F$ is a real number, which corresponds
to an imaginary mass for both the $\sigma$ and the $\pi$ fields. For this
reason one can immediately deduce that the point $\varphi=0$ does not
correspond to the minimum of the potential. Moreover, an expansion around this
point is instable.

The potential $V_{\text{MH}}$ is symmetric under $SO(2)\sim U(1)$ (denoted as
chiral) transformation, namely:
\begin{equation}
\left(
\begin{array}
[c]{c}%
\sigma\\
\pi
\end{array}
\right)  \rightarrow\left(
\begin{array}
[c]{cc}%
\cos\theta & \sin\theta\\
-\sin\theta & \cos\theta
\end{array}
\right)  \left(
\begin{array}
[c]{c}%
\sigma\\
\pi
\end{array}
\right)  \text{ or }\varphi\rightarrow e^{-i\theta}\varphi. \label{chiraltr}%
\end{equation}
The model does not have a unique minimum: all the points $\varphi_{\min
}=Fe^{i\theta}$ for each $\theta\in\lbrack0,2\pi)$ are minima. If no other
information is given, each one of these minima can be in principle realized.
However, we assume that a small perturbation, which breaks chiral symmetry but
does not break parity, $V_{\text{MH}}\rightarrow V_{\text{MH}}-\varepsilon
\sigma$ with $\varepsilon\in0^{+}$, takes place: as a consequence, the only
realized minimum is $\varphi_{\min}=F.$ [A change of sign of $\varepsilon$
would simply provide the equivalent solution $-\varphi_{\min}$]. When
evaluating the fluctuations around the minimum $\varphi_{\min}=F$, one obtains
a scalar, massive $\sigma$ meson with $M_{\sigma}^{2}=2\lambda F^{2}$ and a
pseudoscalar, massless Goldstone boson $\pi$. The chiral symmetry of the model
is not realized as a degeneracy of the particle spectrum because the minimum
(i.e. the vacuum) is not left invariant by this transformation: spontaneous
breaking of chiral symmetry has taken place and the field $\pi$ is the
corresponding Goldstone boson.

\section{QCD origin of the Mexican hat}

For the purpose of this paper we briefly recall how the Mexican hat potential
describes the spontaneous breaking of chiral symmetry which is observed in the
context of low-energy QCD. The matrix $\Phi=S+iP$ includes $N_{f}^{2}$ scalar
and $N_{f}^{2}$ pseudoscalar fields, $S=S^{a}t^{a}$ and $P=P^{a}t^{a}$ where
the matrices $t^{a}$ with $a=1,...,N_{f}^{2}-1$ are the generators of
$SU(N_{f})$ (with $\mathrm{Tr}[t^{a}t^{b}]=\frac{1}{2}\delta^{ab}$) and
$t^{0}=\sqrt{\frac{1}{2N_{f}}}\mathbf{1}_{N_{f}}.$ Upon chiral transformation
$U_{R}(N_{f})\times U_{L}(N_{f})$ the field $\Phi$ transforms as
$\Phi\rightarrow L\Phi R^{\dagger}$ with $L$ $\epsilon$ $U_{L}(N_{f})$ and $R$
$\epsilon$ $U_{R}(N_{f}).$ The transformation in flavor space $SU_{V}(N_{f})$
is obtained by setting $L=R=U_{V},$ where $U_{V}$ is a $SU(N_{f})$ matrix. The
transformation $SU_{A}(N_{f})$ is obtained by setting $L=R^{\dagger}=U_{A}$
$,$ where $U_{A}$ is a $SU(N_{f})$ matrix. (Note, however, that this set of
transformations does not form group for $N_{F}>1$ because two subsequent axial
transformations are not an axial transformation). Finally, the $U_{A}(1)$
axial transformation is obtained by setting $L=R^{\dagger}=e^{-i\alpha
}\mathbf{1}_{N_{f}}$ \cite{fnsigma}. (The $U_{V}(1)$ transformation
corresponds to $L=R=e^{i\alpha}\mathbf{1}_{N_{f}},$ thus trivially implying
the identity transformation $\Phi\rightarrow\Phi$).

The effective potential for the field $\Phi$ reads \cite{GG}
\begin{equation}
V_{\text{eff}}[\Phi;\mu^{2},\gamma,\delta,k,h]=\mathrm{Tr}\left[  \mu^{2}%
\Phi^{\dagger}\Phi+\gamma\left(  \Phi^{\dagger}\Phi\right)  ^{2}\right]
+\delta\left(  \mathrm{Tr}[\Phi^{\dagger}\Phi]\right)  ^{2}-k\,(\det
\Phi^{\dagger}+\det\Phi)-\,\mathrm{Tr}[h(\Phi^{\dagger}+\Phi)]%
\begin{array}
[c]{c}%
.
\end{array}
\label{veff}%
\end{equation}
The first three terms are invariant upon $U_{R}(N_{f})\times U_{L}(N_{f})$
transformations. A sufficient condition for the stability of the potential is
that $\gamma>0$ and $\delta>0.$ The term proportional to $k$ is not invariant
under the $U_{A}(1)$ axial transformation and describes the so-called axial
anomaly \cite{kterm}. In the last term the diagonal $N_{f}\times N_{f}$ matrix
$h$ describes the explicit contribution of nonzero current quark masses. It is
not invariant under $SU_{A}(N_{f})$ and $U_{A}(1)$ transformations, and if
$h\neq$const$\cdot\mathbf{1}_{N_{f}}$, it is also not invariant under
$SU_{V}(N_{f})$.

A first, naive attempt to obtain the Mexican hat of Eq. (\ref{sigma}) is to
study the case $N_{f}=1$ with $\Phi=\sqrt{\frac{1}{2}}(\sigma+i\pi
)=\sqrt{\frac{1}{2}}\varphi.$ In the chiral limit $(h=0)$ one can easily
identify $\lambda=(\gamma+\delta)$ and $\mu^{2}=-(\gamma+\delta)F^{2}<0$. The
latter is a necessary condition for spontaneous symmetry breaking. However,
the anomalous term $-k\,(\det\Phi^{\dagger}+\det\Phi)=-\sqrt{2}k\sigma$ breaks
explicitly chiral symmetry and cannot be regarded as a small perturbation.
This is due to the fact that for $N_{f}=1$ the chiral transformation
$SU_{A}(N_{f})$ cannot be distinguished from the axial transformation
$U_{A}(1).$ We conclude that, in virtue of the anomaly, the Mexican hat
potential cannot be reproduced in the case of one quark flavor only.

When $N_{f}=2$ the matrix $\Phi$ reads
\begin{equation}
\Phi=\sum_{a=0}^{3}\phi_{a}t_{a}=(\sigma+i\eta)\,t^{0}+(\vec{a}_{0}+i\vec{\pi
})\cdot\vec{t}\;, \label{scalars}%
\end{equation}
where $\vec{t}=\vec{\tau}/2,$ with the vector of Pauli matrices $\vec{\tau}$,
and $t^{0}=\mathbf{1}_{2}/2.$

In terms of quark degrees of freedom, the scalar isotriplet $\vec{a}_{0}$ and
the pseudoscalar pion $\vec{\pi}$ are given by $\overline{u}d,$ $\sqrt
{\frac{1}{2}}(\overline{u}u-\overline{d}d),$ $\overline{d}u,$ while the
$\sigma$ and the $\eta$ mesons by$\sqrt{\frac{1}{2}}(\overline{u}%
u+\overline{d}d).$ The identification of the pion triplet with the
experimentally very well known resonances $\pi^{\pm}(139)$ and $\pi^{0}(135)$
listed in the Particle data Group (PDG) \cite{pdg} is straightforward. In the
pseudoscalar-isoscalar channel, one has in Ref. \cite{pdg} two resonances
$\eta(547)$ and $\eta(958),$ which are a combination of the bare contributions
$\eta\equiv\sqrt{\frac{1}{2}}(\overline{u}u+\overline{d}d)$ entering in Eq.
(\ref{scalars}) and the $s$-quark counterpart $\overline{s}s$. The physical
field $\eta(547)$ reads $\eta(547)=\cos(\varphi_{P})\sqrt{\frac{1}{2}%
}(\overline{u}u+\overline{d}d)+\sin(\varphi_{P})\overline{s}s$ where
$\varphi_{P}\simeq-35^{\circ}$ \cite{frascati}, while $\eta(958)$ is the
corresponding orthogonal combination. (One can also `unmix' the two physical
fields and obtain that, in an hypothetical $N_{f}=2$ world without $s$ quark,
the $\eta\equiv\sqrt{\frac{1}{2}}(\overline{u}u+\overline{d}d)$ would have a
mass of about 700 MeV \cite{frascati}). The identification of the fields
$\sigma$ and $\overrightarrow{a}_{0}$ is more complicated and addresses the
problem of the identification of scalar mesons in low-energy QCD. Two set of
candidates are the resonances $\{f_{0}(600),a_{0}(980)\}$ and $\{f_{0}%
(1370),a_{0}(1450)\}.$ A detailed description of this issue is not relevant
for the scope of this paper, see however Ref. \cite{susanna} and refs. therein.

We assume that the charged fields $\pi^{1},$ $\pi^{2},$ $a_{0}^{1},$
$a_{0}^{2}$ do not condense. In this case they are not relevant in the study
of the minima of the potential and we set their mean value to zero. We are
therefore left with the diagonal matrix%
\begin{equation}
\Phi=\frac{1}{2}\left(
\begin{array}
[c]{cc}%
\sigma+a_{0}+i(\eta+\pi) & 0\\
0 & \sigma-a_{0}+i(\eta-\pi)
\end{array}
\right)  \label{philim}%
\end{equation}
where $a_{0}$ and $\pi$ refer to the neutral $a_{0}^{3}$ and $\pi^{3}$ mesons.

The anomaly term of the potential reads explicitly in the case $N_{f}=2$%
\begin{equation}
-k\,(\det\Phi^{\dagger}+\det\Phi)=-\frac{k}{2}(\sigma^{2}+\pi^{2})+\frac{k}%
{2}(a_{0}^{2}+\eta^{2})\text{ }.
\end{equation}
For the case $k>0$, the absolute minimum is found for a nonzero expectation
value of the field $\sigma$ (or $\pi$) and not for a nonzero value of $\eta$
(or $a_{0}$). This is thus the physically interesting case because a
condensation of $\eta$ (or $a_{0}$) would imply a parity (or isospin) breaking
which is not observed in the processes listed in the PDG \cite{pdg}.

By further setting $a_{0}=\eta=0$ the potential (\ref{veff}) reduces exactly
to Eq. (\ref{sigma}) by identifying
\begin{equation}
\lambda=\left(  \frac{\gamma}{2}+\delta\right)
\begin{array}
[c]{c}%
,
\end{array}
\text{ }\mu^{2}=-\left(  \frac{\gamma}{2}+\delta\right)  F^{2}+k%
\begin{array}
[c]{c}%
,
\end{array}
\varepsilon=0%
\begin{array}
[c]{c}%
.
\end{array}
\end{equation}
Note, the choice $\varepsilon=0$ is denoted as the chiral limit. The explicit
inclusion of a breaking term proportional to $h=\varepsilon\mathbf{1}_{2}%
\neq0$ plays the role of the small external perturbation, which induces the
condensation of the $\sigma$ field and not of $\pi$. By further setting the
mean value of $\pi$ to zero, the potential in terms of the field $\sigma$ only
reads%
\begin{equation}
V(\sigma)=\frac{1}{2}\left(  \mu^{2}-k\right)  \sigma^{2}+\frac{1}{4}\left(
\frac{\gamma}{2}+\delta\right)  \sigma^{4}-\varepsilon\sigma\text{ .}%
\end{equation}
The minimum of the latter is realized by a nonzero value $\sigma=\phi\neq0$ if
the quantity $\mu^{2}-k$ is a negative number (at zeroth order in
$\varepsilon$ one has $\phi=F$). In this case spontaneous breaking of chiral
symmetry takes place and the pions emerge as (quasi) Goldstone bosons.

The masses of all the fields, as calculated from Eq. (\ref{veff}) as second
derivatives around the minimum $\sigma=\phi\neq0$, read:%
\begin{align}
M_{\pi}^{2}  &  =\mu^{2}-k+\left(  \frac{\gamma}{2}+\delta\right)  \phi
^{2}=\frac{\varepsilon}{\phi},\text{ }M_{\eta}^{2}=\mu^{2}+k+\left(
\frac{\gamma}{2}+\delta\right)  \phi^{2}\\
M_{\sigma}^{2}  &  =\mu^{2}-k+3\left(  \frac{\gamma}{2}+\delta\right)
\phi^{2},\text{ }M_{a_{0}}^{2}=\mu^{2}+k+\left(  \frac{3}{2}\gamma
+\delta\right)  \phi^{2}%
\end{align}
It is clear that $M_{\eta}^{2}$ receive a positive contribution form the
anomalous term $k>0;$ this also explains while the latter is clearly heavier
than the pion fields. It is also renowned that the axial current reads
$J_{A,\mu}^{a}=\phi\partial_{\mu}\pi^{a}$: the constant $\phi$ can then be set
equal to the pion decay constant $f_{\pi}=92.4$ MeV (details, for instance, in
Ref. \cite{koch}).

As a last remark we note that, in the limit of Eq. (\ref{philim}), the matrix
$\Phi$ can be written as: $\Phi=\frac{1}{2}diag\{\sigma+i\pi,\sigma
-i\pi\}=\sigma t^{0}+i\pi t^{3}$. Therefore, a $SU_{A}(2)$ transformation
$\Phi\rightarrow U_{A}\Phi U_{A}$ with $U_{A}$ $\epsilon$ $SU_{A}(2)$ in the
third isospin direction, i.e. $U_{A}=e^{-i\alpha t^{3}},$ is such that
$\Phi\rightarrow U_{A}\Phi U_{A}=U_{A}^{2}\Phi=e^{-2i\alpha t^{3}}\Phi.$ The
latter reduces exactly to the transformation of Eq. (\ref{chiraltr}), i.e.
$\varphi=\sigma+i\pi\rightarrow e^{-i\theta}\varphi$ by identifying
$\alpha=\theta/2$. We thus obtain the simple Mexican hat potential in Eq.
(\ref{sigma}) as a special case of the general $N_{f}=2$ effective potential
by identifying $\sigma$ as the scalar-isoscalar field and $\pi$ as the
pseudoscalar neutral member of the isotriplet field $\vec{\pi}$.

\section{Two Mexican hats }

Let us now turn to the case of interest of this work: two linked Mexican hats.
A `double Mexican hat potential' is introduced in terms of the complex fields
$\varphi_{1}=\sigma_{1}+i\pi_{1}$ and $\varphi_{2}=\sigma_{2}+i\pi_{2}$:
\begin{equation}
V_{\text{DMH}}=\frac{\lambda_{1}}{4}\left(  \varphi_{1}^{\ast}\varphi
_{1}-F_{1}^{2}\right)  ^{2}+\frac{\lambda_{2}}{4}\left(  \varphi_{2}^{\ast
}\varphi_{2}-F_{2}^{2}\right)  ^{2}+\frac{c}{2}\left[  \left(  \varphi
_{2}^{\ast}\varphi_{1}\right)  ^{2}+\left(  \varphi_{1}^{\ast}\varphi
_{2}\right)  ^{2}\right]  .\label{vdmh}%
\end{equation}
As long as $F_{1}$ and $F_{2}$ are real numbers, it constitutes of two
distinct Mexican hats for $\varphi_{1}$ and $\varphi_{2},$ and a $c$-term,
which mixes them \cite{fn2}. The fields $\sigma_{1}$ and $\sigma_{2}$ are
assumed to have positive parity, while the fields $\pi_{1}$ and $\pi_{2}$
negative parity.

The model of Eq. (\ref{vdmh}) is manifestly invariant under the
\textquotedblleft chiral\textquotedblright\ $U(1)$ transformation applied to
\emph{both} fields:%
\begin{equation}
\varphi_{1}\rightarrow e^{-i\theta}\varphi_{1},\text{ }\varphi_{2}\rightarrow
e^{-i\theta}\varphi_{2}. \label{u1}%
\end{equation}
The condition $\lambda_{1},\lambda_{2}>0$ is obviously necessary to guarantee
the stability of the potential. Simple algebra shows that a further constraint
is needed: the parameter $c$ must be such that $\left\vert c\right\vert
<\min\{\frac{\lambda_{1}+\lambda_{2}}{4},\frac{\sqrt{\lambda_{1}\lambda_{2}}%
}{2}\}.$ We also set, for definiteness, $F_{1}<F_{2}.$

Note that if $c=0$ the model reduces to two decoupled linear sigma models. The
symmetry is in this limit larger: $U^{(1)}(1)\times U^{(2)}(1),$ i.e. it is
invariant under $\varphi_{1}\rightarrow e^{-i\theta_{1}}\varphi_{1}$ or
$\varphi_{2}\rightarrow e^{-i\theta_{2}}\varphi_{2}$ separately. Two Goldstone
bosons $\pi_{1}$ and $\pi_{2}$ and two massive $\sigma_{1}$ and $\sigma_{2}$
fields with $M_{\sigma_{1}}^{2}=2\lambda_{1}F_{1}^{2}$ and $M_{\sigma_{2}}%
^{2}=2\lambda_{2}F_{2}^{2}$ are obtained.

In terms of the fields $(\sigma_{1},\pi_{1})$ and $(\sigma_{2},\pi_{2})$ the
potential $V_{DMH}$ takes the form%
\begin{equation}
V_{\text{DMH}}=\frac{\lambda_{1}}{4}\left(  \sigma_{1}^{2}+\pi_{1}^{2}%
-F_{1}^{2}\right)  ^{2}+\frac{\lambda_{2}}{4}\left(  \sigma_{2}^{2}+\pi
_{2}^{2}-F_{2}^{2}\right)  ^{2}+c\left[  \left(  \sigma_{1}^{2}-\pi_{1}%
^{2}\right)  \left(  \sigma_{2}^{2}-\pi_{2}^{2}\right)  +4\sigma_{1}\pi
_{1}\sigma_{2}\pi_{2}\right]  .
\end{equation}
As usual, the minima of the model must be identified. The sign of the
parameter $c$ plays an important role: the cases $c\leq0$ and $c>0$ are
studied separately later on, after that we have related the potential to a
generalized hadronic model.

As studied above in the presence of only one complex scalar field $\varphi,$
the potential $V_{\text{DMH}}$ may arise as a special case of a more general
$N_{f}=2$ QCD effective theory in which one starts from two matrices $\Phi
_{1}$ and $\Phi_{2},$ each one made of $N_{f}^{2}$ scalar and $N_{f}^{2}$
pseudoscalar fields as in Eq. (\ref{scalars}). The matrix $\Phi_{1}$
represents the ground state (pseudo)scalar fields, while $\Phi_{2}$ the first
radial excitation. The effective potential reads
\begin{equation}
V_{\text{eff}}[\Phi_{1},\Phi_{2}]=V_{\text{eff}}^{(1)}[\Phi_{1}]+V_{\text{eff}%
}^{(2)}[\Phi_{2}]+2c\mathrm{Tr}\left[  (\Phi_{2}^{\dagger}\Phi_{1})^{2}%
+(\Phi_{1}^{\dagger}\Phi_{2})^{2}\right]
\begin{array}
[c]{c}%
,
\end{array}
\label{veffgen}%
\end{equation}
where $V_{\text{eff}}^{(1)}[\Phi_{1}]$ and $V_{\text{eff}}^{(2)}[\Phi_{2}]$
read as in Eq. (\ref{veff}):
\begin{equation}
V_{\text{eff}}^{(1)}[\Phi_{1}]=V_{\text{eff}}[\Phi_{1};\mu_{1}^{2},\gamma
_{1},\delta_{1},k_{1},h_{1}=\varepsilon_{1}\mathbf{1}_{2}]%
\begin{array}
[c]{c}%
,
\end{array}
V_{\text{eff}}^{(2)}[\Phi_{1}]=V_{\text{eff}}[\Phi_{2};\mu_{2}^{2},\gamma
_{2},\delta_{2},k_{2},h_{2}=\varepsilon_{2}\mathbf{1}_{2}]%
\begin{array}
[c]{c}%
.
\end{array}
\end{equation}
The $U_{R}(N_{f})\times U_{L}(N_{f})$ chiral transformation implies the
simultaneous transformation of both fields
\begin{equation}
\Phi_{1}\rightarrow L\Phi_{1}R^{\dagger}%
\begin{array}
[c]{c}%
,
\end{array}
\Phi_{2}\rightarrow L\Phi_{2}R^{\dagger}%
\begin{array}
[c]{c}%
.
\end{array}
\end{equation}
By performing the same steps as before, we reduce the matrices $\Phi_{1(2)}$
to their diagonal form $\Phi_{1(2)}=\frac{1}{2}diag\{\sigma_{1(2)}+i\pi
_{1(2)},\sigma_{1(2)}-i\pi_{1(2)}\}$. A $SU_{A}(2)$ chiral transformation in
the third isospind direction reduces to Eq. (\ref{u1}). The identification of
the parameters of Eq. (\ref{vdmh}) with those of Eq. (\ref{veffgen}) leads to%
\begin{align}
\lambda_{1}  &  =\left(  \frac{\gamma_{1}}{2}+\delta_{1}\right)
\begin{array}
[c]{c}%
,
\end{array}
\mu_{1}^{2}-k_{1}=-\left(  \frac{\gamma_{1}}{2}+\delta_{1}\right)  F_{1}^{2}%
\begin{array}
[c]{c}%
,
\end{array}
\varepsilon_{1}=0%
\begin{array}
[c]{c}%
,
\end{array}
\\
\lambda_{2}  &  =\left(  \frac{\gamma_{2}}{2}+\delta_{2}\right)
\begin{array}
[c]{c}%
,
\end{array}
\mu_{2}^{2}-k_{2}=-\left(  \frac{\gamma_{2}}{2}+\delta_{2}\right)  F_{2}^{2}%
\begin{array}
[c]{c}%
,
\end{array}
\varepsilon_{2}=0%
\begin{array}
[c]{c}%
.
\end{array}
\end{align}
Two Mexican hats are present as long as $F_{1}$ and $F_{2}$ are real numbers,
i.e. if the quantities $\mu_{1}^{2}-k_{1}$ and $\mu_{2}^{2}-k_{2}$ are
negative real numbers. In this case, one has a Mexican hat for $V_{\text{eff}%
}[\Phi_{1}=\frac{1}{2}diag\{\sigma_{1}+i\pi_{1},\sigma_{1}-i\pi_{1}\},\Phi
_{2}=0]$ (in the subspace of the ground state fields $\{\sigma_{1},\pi_{1}\}$)
and also for $V_{\text{eff}}[\Phi_{1}=0,\Phi_{2}=\frac{1}{2}diag\{\sigma
_{2}+i\pi_{2},\sigma_{2}-i\pi_{2}\}]$ (in the subspace of $\{\sigma_{2}%
,\pi_{2}\}$).

Note, in Ref. \cite{cohen} a Lagrangian with (an infinity of) linked $\Phi
_{k}$ has been introduced, but only one Mexican hat is present: while $\mu
_{1}^{2}-k_{1}<0,$ one has $\mu_{p}^{2}-k_{p}>0$ for $p=2,3,...$ Similarly, in
the $N_{f}=3$ models of Refs. \cite{fariborz} an additional nonet of scalar
and pseudoscalar tetraquark mesons is introduced, but the Mexican hat is
present only in the subspace of the ground-state quark-antiquark
(pseudo)scalar mesons. In the recent work of Ref. \cite{espriu} two multiplets
$\Phi_{1}$ and $\Phi_{2}$ are considered in a general fashion, but the
attention is focused on parity breaking at nonzero temperatures/densities.

More in general, we also refer to Higgs sector of supersymmetric models (Ref.
\cite{ss} and refs. therein) and to works on superconductivity (Refs.
\cite{solid} and refs. therein) where scalar theories, their mixing and
spontaneous symmetry breaking are studied.

\section{Condensation with no spontaneous breaking of parity }

We study the minima of the potential $V_{\text{DMH}}$ of Eq. (\ref{vdmh}) for
$-c_{\max}<c\leq0.$ One absolute minimum of the potential $V_{\text{DMH}}$ is
given by%
\begin{equation}
\left(  \pi_{1}=\pi_{2}=0,\text{ }\sigma_{1}=A_{1},\text{ }\sigma_{2}%
=A_{2}\right)  \leftrightarrow\left(  \varphi_{1}=A_{1},\varphi_{2}%
=A_{2}\right)
\begin{array}
[c]{c}%
,
\end{array}
\label{mincmin0}%
\end{equation}

\begin{equation}
A_{1}=\sqrt{\frac{F_{1}^{2}-\frac{2c}{\lambda_{1}}F_{2}^{2}}{1-\frac{4c^{2}%
}{\lambda_{1}\lambda_{2}}}}%
\begin{array}
[c]{c}%
,
\end{array}
A_{2}=\sqrt{\frac{F_{2}^{2}-\frac{2c}{\lambda_{2}}F_{1}^{2}}{1-\frac{4c^{2}%
}{\lambda_{1}\lambda_{2}}}}%
\begin{array}
[c]{c}%
.
\end{array}
\end{equation}
Due to the form of the potential this minimum is not unique. All other minima
can be obtained by applying a chiral $U(1)$ transformation to Eq.
(\ref{mincmin0}):%
\begin{equation}
\left(  \varphi_{1,\min},\varphi_{2,\min}\right)  =\left(  A_{1}e^{i\theta
},A_{2}e^{i\theta}\right)  \text{ with }\theta\in\lbrack0,2\pi)%
\begin{array}
[c]{c}%
.
\end{array}
\end{equation}
The minimum of Eq. (\ref{mincmin0}) is unequivocally realized if we add to the
potential the following parity-conserving but chirally breaking terms%
\begin{equation}
V_{\text{DMH}}\rightarrow V_{\text{DMH}}-\varepsilon_{1}\sigma_{1}%
-\varepsilon_{2}\sigma_{2}\text{ with }\varepsilon_{1},\varepsilon_{2}\in0^{+}%
\begin{array}
[c]{c}%
.
\end{array}
\end{equation}
Note, the latter shift corresponds to small but nonzero current quark masses,
$h_{1}=$ $\varepsilon_{1}\mathbf{1}_{2}$, $h_{2}=$ $\varepsilon_{2}%
\mathbf{1}_{2},$ in Eq. (\ref{veffgen}).

Clearly, the minimum of Eq. (\ref{mincmin0}) is parity-conserving because two
scalar fields condense. Being not invariant under chiral transformation, a
spontaneous breaking of this symmetry occurs in the vacuum. The behavior of
the condensates as function of the parameter $c$ is reported in Fig. 1, left
panel $(c\leq0)$ for a paradigmatic numerical choice.

The mass matrices in both the scalar and the pseudoscalar sectors are obtained
by calculating second-order derivatives evaluated at the point given in Eq.
(\ref{mincmin0}). They explicitly read \cite{fn3}:%
\begin{equation}
\left(
\begin{array}
[c]{cc}%
M_{\sigma_{1}}^{2}=3\lambda_{1}A_{1}^{2}-\lambda_{1}F_{1}^{2}+2cA_{2}^{2} &
4cA_{1}A_{2}\\
4cA_{1}A_{2} & M_{\sigma_{2}}^{2}=3\lambda_{2}A_{2}^{2}-\lambda_{2}F_{2}%
^{2}+2cA_{1}^{2}%
\end{array}
\right)  ;
\end{equation}%
\begin{equation}
\left(
\begin{array}
[c]{cc}%
M_{\pi_{1}}^{2}=\lambda_{1}(A_{1}^{2}-F_{1}^{2})-2cA_{2}^{2} & 4cA_{1}A_{2}\\
4cA_{1}A_{2} & M_{\pi_{2}}^{2}=\lambda_{2}(A_{2}^{2}-F_{2}^{2})+2cA_{1}^{2}%
\end{array}
\right)  . \label{psmat}%
\end{equation}
The `physical masses' $M_{\sigma_{1}^{\prime}},$ $M_{\sigma_{2}^{\prime}},$
$M_{\pi_{1}^{\prime}},$ $M_{\pi_{2}^{\prime}}$ (the first two states with
positive parity, the latter two with negative parity) are obtained in the
standard way as eigenvalues of the latter two matrices. The spectrum of the
system consists of two massive scalar fields, one massive pseudoscalar field
and one massless pseudoscalar Goldstone boson. In fact, one eigenvalue of the
pseudoscalar matrix of Eq. (\ref{psmat}) vanishes, therefore realizing the
Goldstone theorem. In Fig. 1, right panel, the masses are plotted as function
of $c<0$ for a particular numerical choice. Obviously, no mass degeneracy is
present due to the fact that chiral symmetry is spontaneously broken. Notice
also that the mass of the massive pseudoscalar meson $M_{\pi_{2}^{\prime}}$
vanishes for $c\rightarrow0^{-},$ in agreement with the fact that a second
Goldstone boson exists due to the larger, spontaneously broken symmetry in
this limit.%
\begin{figure}
[ptb]
\begin{center}
\includegraphics[
height=2.0211in,
width=6.4057in
]%
{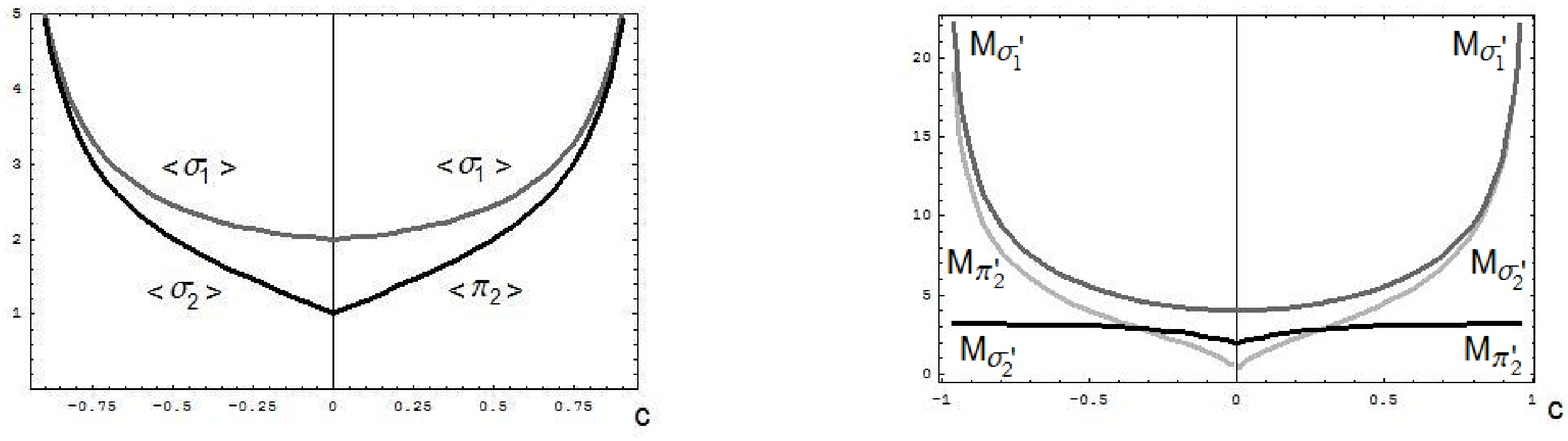}%
\caption{Fixing $\lambda_{1}=\lambda_{2}=1,$ $F_{2}=F_{1}/2=1$ (unit energy),
the condensates corresponding to Eq. (\ref{mincmin0}) for $-1=-c_{\max}<c<0$
and to Eq. (\ref{mincmag0}) for $0<c<c_{\max}$ are plotted in the left panel.
For $c<0$ parity is conserved, for $c>0$ is spontaneously broken. In the right
panel the masses of the three physical massive states are plotted: (the state
$\pi_{1}^{\prime}$ is the massless Goldstone boson for each $c$). For $c<0$
the states have definite parity, for $c>0$ no state is a parity eigenstate.}%
\end{center}
\end{figure}

Some considerations are in order:

(i) If the parameter $F_{2}$ -instead of being real- is a purely imaginary
number (i.e. if $\mu_{2}^{2}-k_{2}>0$) the model has different properties:
only one Mexican hat in the subspace of $\sigma_{1}$ and $\pi_{1}$ is present.
As a consequence, only the field $\sigma_{1}$ condenses to $F_{1}$ (chiral
condensate) and $\pi_{1}$ is the Goldstone boson \cite{fnint}. Denoting
$F_{2}=i\alpha$ one obtains: $M_{\sigma_{1}}^{2}=2\lambda_{1}F_{1}^{2},$
$M_{\pi_{1}}=0,$ $M_{\sigma_{2}}^{2}=\lambda_{2}\alpha^{2}+cF_{1}^{2}$ and
$M_{\pi_{2}}^{2}=\lambda_{2}\alpha^{2}-cF_{1}^{2}.$ The mass splitting between
$\sigma_{2}$ and $\pi_{2}$ is generated by the chiral condensate $\sigma
_{1}=F_{1}.$ This is the typical picture for low-energy QCD effective
theories, in which the fields $\sigma_{2},$ $\pi_{2}$ are interpreted as the
radial excitations of the ground state $\sigma_{1},$ $\pi_{1}$ \cite{cohen}.
If, for heavier multiplets $\Phi_{k}$, one has smaller and smaller $c$, one
recovers the degeneracy of the chiral partners. For a more detailed
description of chiral symmetry restoration see Ref. \cite{fnint} and refs. therein.

(ii) The scenario of two Mexican hats together with $c<0$ \emph{cannot} be
excluded as an effective theory of QCD. Although a phenomenological study in
the framework of a realistic potential should be performed to investigate this
possibility, here we simply note that the case with two Mexican hats (with
$c<0$) is in agreement with all the symmetries and constraints imposed by QCD.

(iii) The case $c=0$ is interesting. It implies that a larger symmetry group
is realized for the effective theory than at the fundamental level. In fact,
in this limit the effective theory of Eq. (\ref{veffgen}) is invariant under
$\Phi_{1}\rightarrow L_{1}\Phi_{1}R_{1}^{\dagger}$ and, independently, under
$\Phi_{2}\rightarrow L_{2}\Phi_{2}R_{2}^{\dagger}$, i.e. under the product of
independent chiral transformations $U_{R}^{(1)}(N_{f})\times U_{L}^{(1)}%
(N_{f})\times U_{R}^{(2)}(N_{f})\times U_{L}^{(2)}(N_{f})$. (The latter
transformations reduce in the toy model to the already mentioned invariance
under $U^{(1)}(1)\times U^{(2)}(1),$ i.e.$\varphi_{1}\rightarrow
e^{-i\theta_{1}}\varphi_{1}$ and, independently, $\varphi_{2}\rightarrow
e^{-i\theta_{2}}\varphi_{2}$ when axial transformations in the third isospin
direction are considered.) If $F_{2}$ is a real number, this would imply the
presence of two Goldstone bosons, an eventuality which is not seen in the real
world. Indeed, the parameter $c$ should also not be too small, otherwise a
second, light pseudoscalar meson would be present in the spectrum, see Fig. 1,
right panel, what is excluded by experimental data (the second pionic
excitation has a mass of about $1.3$ GeV \cite{pdg}). If $F_{2}$ is imaginary
as described in the point (ii), the condition $c=0$ implies the degeneracy
$M_{\sigma_{2}}^{2}=\lambda_{2}\alpha^{2}+cF_{1}^{2}$ and $M_{\pi_{2}}%
^{2}=\lambda_{2}\alpha^{2}-cF_{1}^{2}.$ In the context of the already
mentioned effective restoration of chiral symmetry, where for heavier
multiplets a degeneracy is postulated, one indeed would have an approximately
higher symmetry, corresponding to a product of $U_{R}^{(k)}(N_{f})\times
U_{L}^{(k)}(N_{f})$ for different values of $k$, where $k$ refers to the
$k$-th excited (pseudo)scalar matrix $\Phi_{k}.$

(iv) A generalization to more than 2 Mexican hats can also be easily
performed. However, in order to avoid a proliferation of undesired light
pseudoscalar mesons, the mixing among the different $\Phi_{k}$ should be
large. We regard this possibility as remote for QCD, see next point.

(v) QCD in the chiral limit has only dimensional parameter, the Yang-Mills
scale $\Lambda_{QCD}.$ By varying it, it is -although speculative- conceivable
that different phases are realized: a phase in which no Mexican hat is present
($F_{1}$ and $F_{2}$ both purely imaginary, with no spontaneous chiral
symmetry breaking and no Goldstone boson(s)) obtained for $0<\Lambda_{QCD}%
\leq\Lambda_{1}$ \cite{casher}, a phase in which only for the ground state
mesons one has a Mexican hat ($F_{1}$ real and $F_{2}$ purely imaginary, which
is the standard scenario) for $\Lambda_{1}\leq\Lambda_{QCD}\leq\Lambda_{2}$, a
phase in which two Mexican hats are present $(F_{1}$ and $F_{2}$ both real)
for $\Lambda_{2}\leq\Lambda_{QCD}\leq\Lambda_{3},$ and so on and so forth. The
case $\Lambda_{2}\leq\Lambda_{QCD}\leq\Lambda_{3}$ is the one described by the
potential of Eq. (\ref{vdmh}) when both $F_{1}$ and $F_{2}$ are real numbers
\cite{fnlamda}.

(vi) In the case of a double Mexican potential ($F_{1}$ and $F_{2}$ real), it
is not possible to obtain a simple situation as described by the potential
(\ref{veff}), or its reduced form (\ref{sigma}). In the scenario of two
Mexican hats it is therefore necessary to take into account both multiplets
$\Phi_{1}$ and $\Phi_{2}.$ More in general, in the presence of more Mexican
hats, one is obliged to include all of them in a linear hadronic theory of
QCD. Needles to say, a double (or multiple) Mexican hat would correspond to a
substantial complication. `Life is easier' if such a scenario is not realized
and if only the ground state $\sigma\equiv\sigma_{1}$ condenses. Nevertheless,
the question why this should be the case is interesting. Is there some yet
unknown motivations which forbids the emergence of a second (or more) Mexican
hat(s)? Can it be an accidental fact, which depends only on the value of
$\Lambda_{QCD}$ as describe above?

\section{Condensation with spontaneous breaking of parity\emph{ }}

We now study\emph{ }$V_{\text{DMH}}$ for $0<c<c_{\max}.$ One absolute minimum
is given by%
\begin{equation}
\left(  \pi_{1}=\sigma_{2}=0,\text{ }\sigma_{1}=B_{1},\text{ }\pi_{2}%
=B_{2}\right)  \leftrightarrow\left(  \varphi_{1}=B_{1},\varphi_{2}%
=B_{2}e^{i\pi/2}\right)  ,\label{mincmag0}%
\end{equation}

\begin{equation}
B_{1}=\sqrt{\frac{F_{1}^{2}+\frac{2c}{\lambda_{1}}F_{2}^{2}}{1-\frac{4c^{2}%
}{\lambda_{1}\lambda_{2}}}},\text{ }B_{2}=\sqrt{\frac{F_{2}^{2}+\frac
{2c}{\lambda_{1}}F_{1}^{2}}{1-\frac{4c^{2}}{\lambda_{1}\lambda_{2}}}}.
\end{equation}
The pseudoscalar field $\pi_{2}$ assumes a nonzero vacuum expectation value.
This minimum is not unique: the full set of minima is obtained by performing a
$U(1)$ rotation of Eq. (\ref{mincmag0}):
\begin{equation}
(\varphi_{1}=B_{1}e^{i\theta},\varphi_{2}=B_{2}e^{i(\theta+\pi/2)})\text{ with
}0\leq\theta<2\pi.
\end{equation}
Each of these minima breaks parity because $\pi_{1}$ and $\pi_{2}$ \emph{never
}vanish simultaneously. By adding to the system the parity conserving but
chirally breaking term $V_{DMH}\rightarrow V_{DMH}-\varepsilon_{1}\sigma
_{1}-\varepsilon_{2}\sigma_{2},$ Eq. (\ref{mincmag0}) is the univocally
selected minimum: in fact, this is the point at which $\sigma_{1}$ is maximal
for the assumed ordering $F_{1}<F_{2}$. Note that, although a parity
conserving perturbation has been added, still the realized vacuum breaks
parity. We conclude that in the proposed model, besides spontaneous breaking
of chiral symmetry, also a spontaneous breaking of parity takes place in the
vacuum for $c>0.$ In Fig. 1, left panel, the condensate of Eq. (\ref{mincmag0}%
) are plotted for $c>0$.

The determination of the physical masses is obtained in the standard fashion.
The crucial difference with respect to the case $c<0$ is that the states of
opposite parity $\sigma_{1}$ and $\pi_{2}$ mix, thus originating two massive
physical states $\sigma_{1}^{\prime}$ and $\pi_{2}^{\prime}$ which are not
eigenstates of parity. At the same time also the states of opposite parity
$\sigma_{2}$ and $\pi_{1}$ mix, out of which one massless and one massive
bosons $\pi_{1}^{\prime}$ and $\sigma_{2}^{\prime}$ -both with undefined
parity- are obtained. Numerically, one has a mirror-like picture for $c>0$
with respect to the parity conserving case, as depicted in Fig. 1, right panel.

Obviously, the here outlined scenario for $c>0$ cannot describe QCD, where
parity is conserved in the vacuum. The Vafa-Witten theorem \cite{vw} states
that spontaneous parity violation does not occur in theories containing
vector-like fermions. Thus, if this theorem holds, the model of Eq.
(\ref{vdmh}) with $c>0$ \emph{cannot} be an effective description of QCD even
when varying $\Lambda_{QCD}$: it is not possible that $F_{1}$ and $F_{2}$ are
real numbers and that at the same time $c$ is negative. However, the validity
of the Vafa-Witten theorem has been questioned in a variety of works (see the
discussion in Ref. \cite{einhorn} and refs. therein). If it is not valid, it
is still conceivable that for a different value of $\Lambda_{QCD}$,
spontaneous breaking of parity takes place in the vacuum: in this case the
here outlined model -with real $F_{2}$ and negative $c$- would correspond to
its low-energy hadronic (confined) realization.

More in general, the original constrain that the charged components of the
pion field can also be released. All the present treatment is still valid upon
replacing $\pi_{2}$ with $\left\vert \overrightarrow{\pi}\right\vert .$ We
have in this case the condition $\left\vert \overrightarrow{\pi}\right\vert
=B_{2}$: as soon as also $\pi^{1}\neq0$ and/or $\pi^{2}\neq0$ not only parity,
but also charge conjugation is spontaneously broken. However, it is enough
that a further, small perturbation, which originates from other interactions
and is invariant under change conjugation, is present: then this additional
perturbation generates a condensation of $\pi^{3}\equiv\pi^{0}$ only, in line
with the discussion of the present paper.

\section{Conclusions}

The main interest of this paper has been the possibility that an hadronic,
$\sigma$-model for QCD is effectively described by a `double' Mexican hat
effective potential. In this scenario not only in the subspace of the neutral
ground state (pseudo)scalar mesons $\sigma\equiv\sigma_{1}$ and $\pi\equiv
\pi_{1}$ fields, but also in the subspace of the first excited (pseudo)scalar
mesons $\sigma\equiv\sigma_{2}$ and $\pi\equiv\pi_{2}$ fields, a typical
Mexican hat form is present. Mixing among these bare configurations arise: in
the case that no spontaneous parity breaking occurs (here for $c<0)$ the
outlined effective model is in agreement with all the constraints imposed by
QCD. In the case that parity symmetry breaking occurs $(c>0)$ the described
model can provide an effective description of a underlying QCD-like theory
only if the Vafa-Witten theorem does not strictly hold. More in general, the
here presented model can also be conceived as an `elementary' model of
(pseudo)scalar fields which generates parity breaking for some choices of the
parameters and may play a role in the early Universe.

\bigskip

\textbf{Acknowledgments: }the author thanks T. Brauner and D. H. Rischke for
useful discussions.


\begin{thebibliography}{99}                                                                                               %


\bibitem {zee}A.~Zee, \textquotedblleft Quantum field theory in a
nutshell,\textquotedblright\
\textit{Princeton, UK: Princeton Univ. Pr. (2003) 518 p}


\bibitem {fnsigma}In the $N_{f}=2$ case one can consider the reduced
combination$\ \Sigma=\sigma t^{0}+i$ $\overrightarrow{\pi}\overrightarrow{t},$
which also transforms as $\Sigma\rightarrow R\Sigma L^{\dagger}$ under chiral
$SU_{R}(2)\times SU_{L}(2)$ transformation (but not under $U_{R}(2)\times
SU_{L}(2)$: the axial transformation, which mixes $\sigma$ with $\eta$ and
$\overrightarrow{\pi}$ with $\overrightarrow{a}_{0}$ cannot be described with
$\Sigma$ only). The contributions of the $\gamma$ and $\delta$ terms are equal
in this case and the corresponding potential can be written as $V_{MH}%
=\frac{\lambda}{4}\left(  2Tr\left[  \Sigma^{\dagger}\Sigma\right]
-F^{2}\right)  ^{2}=\frac{\lambda}{4}\left(  \sigma^{2}+\overrightarrow{\pi
}^{2}-F^{2}\right)  ^{2}.$ Clearly, the model of Eq. (\ref{sigma}) is
recovered by setting $\pi^{1}=\pi^{2}=0.$

\bibitem {GG}S.~Gasiorowicz and D.~A.~Geffen,
Rev.\ Mod.\ Phys.\ \textbf{41}, 531 (1969).


\bibitem {kterm}The anomaly term proportional to $k$ is in the present form
not renormalizable for $N_{f}\geq5.$ Although this is already in the region of
heavy quarks and therefore unimportant for practical purposes, the general
issue is if non-renormalizable terms should enter in an effective description
of QCD. In principle, being an effective QCD model valid in a restricted
low-energy domain, there is no reason to disregard non-renormalizable terms.
For instance, Nambu Jona-Lasinio inspired models of QCD are
non-renormalizable. As another example, one can consider the
non-renormalizable Fermi Lagrangian of weak interactions, which arises as a
low-energy effective term of the more complete (and renormalizable)
electroweak Lagrangian. On the other hand, what can constrain the
dimensionality of terms entering in an hadronic Lagrangian is -rather than the
renormalizability- the requirement that dilatation invariance is solely broken
by a Yang-Mills scale in the glueball sector, see details in Ref.
F.~Giacosa,
arXiv:0903.4481 [hep-ph]
and in Ref.\cite{susanna}.


\bibitem {pdg}C. Amsler et al. (Particle Data Group), Physics Letters
\textbf{B667}, 1 (2008)



\bibitem {frascati}T.~Feldmann and P.~Kroll,
Phys.\ Scripta \textbf{T99} (2002) 13 [arXiv:hep-ph/0201044].
F.~Giacosa,
arXiv:0712.0186 [hep-ph].


\bibitem {susanna}S.~Gallas, F.~Giacosa and D.~H.~Rischke,
arXiv:0907.5084 [hep-ph].
D.~Parganlija, F.~Giacosa and D.~H.~Rischke,
PoS C \textbf{ONFINEMENT8} (2008) 070 [arXiv:0812.2183 [hep-ph]].
D.~Parganlija, F.~Giacosa and D.~H.~Rischke,
AIP Conf.\ Proc.\ \textbf{1030} (2008) 160 [arXiv:0804.3949 [hep-ph]].


\bibitem {koch}V.~Koch,
arXiv:nucl-th/9512029.


\bibitem {fn2}Further $U(1)$ symmetric terms could be added to the potential,
such as the terms of order 2 $\left(  \varphi_{1}^{\ast}\varphi_{2}%
+\varphi_{2}^{\ast}\varphi_{1}\right)  $ and of order 4 $\left(  \varphi
_{1}^{\ast}\varphi_{2}+\varphi_{2}^{\ast}\varphi_{1}\right)  \varphi_{1}%
^{\ast}\varphi_{1}$ and $\left(  \varphi_{1}^{\ast}\varphi_{2}+\varphi
_{2}^{\ast}\varphi_{1}\right)  \varphi_{2}^{\ast}\varphi_{2},$ and also the
$U^{(1)}(1)\times U^{(2)}(1)$ symmetric term $\left(  \varphi_{1}^{\ast
}\varphi_{1}\right)  \left(  \varphi_{2}^{\ast}\varphi_{2}\right)  .$ Here we
restrict for definiteness and simplicity to the term of Eq. (\ref{vdmh}),
leaving an extended study for the future.

\bibitem {cohen}T.~D.~Cohen and L.~Y.~Glozman,
Mod.\ Phys.\ Lett.\ A \textbf{21} (2006) 1939 [arXiv:hep-ph/0512185].




\bibitem {fariborz}A.~H.~Fariborz, R.~Jora and J.~Schechter,
Phys.\ Rev.\ D \textbf{72} (2005) 034001.
M.~Napsuciale and S.~Rodriguez,
Phys.\ Rev.\ D \textbf{70} (2004) 094043, [arXiv:hep-ph/0407037].
F.~Giacosa,
Phys.\ Rev.\ D \textbf{75} (2007) 054007, [arXiv:hep-ph/0611388].


For a simple description of the $N_{f}=2$ case see: A.~Heinz, S.~Struber,
F.~Giacosa and D.~H.~Rischke,
Phys.\ Rev.\ D \textbf{79} (2009) 037502, arXiv:0805.1134 [hep-ph].


\bibitem {espriu}A.~A.~Andrianov, V.~A.~Andrianov and D.~Espriu,
Phys.\ Lett.\ B \textbf{678} (2009) 416 [arXiv:0904.0413 [hep-ph]].


\bibitem {ss}I.~J.~R.~Aitchison,
arXiv:hep-ph/0505105.




\bibitem {solid}E.~Babaev,
arXiv:cond-mat/0302218.
E.~Di Grezia, S.~Esposito and G.~Salesi,
Physics Letters A\textbf{ 373} (2009) 2385 [arXiv:0807.1414
[cond-mat.supr-con]].


\bibitem {fn3}In principle a $4\times4$ matrix should be written down.
However, the entries $\sigma_{1}$-$\pi_{1},$ $\sigma_{1}$-$\pi_{2},...$vanish.

\bibitem {fnint}L.~Y.~Glozman,
Phys.\ Rept.\ \textbf{444} (2007) 1 [arXiv:hep-ph/0701081].

\bibitem {casher}Note, this possibility is not in agreement with the Casher's
argument, which states that in the confining mode chiral symmetry is
necessarily broken, see: A. Casher, Phys. Lett. \textbf{B83}, 395 (1979). If
the Casher's argument is strictly valid, than we are led to conclude that
$\Lambda_{1}=0,$ i.e. a confining but chirally symmetric phase is not possible.

\bibitem {fnlamda}Note, it is assumed that confinement holds for each
$\Lambda_{QCD}.$ Obviously, a direct calculations of the various $\Lambda_{i}$
is not doable at present.

\bibitem {vw}C.~Vafa and E.~Witten,
Phys.\ Rev.\ Lett.\ \textbf{53} (1984) 535.


\bibitem {einhorn}M.~B.~Einhorn and J.~Wudka,
Phys.\ Rev.\ D \textbf{67} (2003) 045004 [arXiv:hep-ph/0205346].

\end{thebibliography}
\end{document}